\PassOptionsToPackage{unicode}{hyperref}
\PassOptionsToPackage{hyphens}{url}
\documentclass[
  11pt,
]{article}
\usepackage{lmodern}
\usepackage{amssymb,amsmath}
\usepackage{ifxetex,ifluatex}
\ifnum 0\ifxetex 1\fi\ifluatex 1\fi=0 
  \usepackage[T1]{fontenc}
  \usepackage[utf8]{inputenc}
  \usepackage{textcomp} 
\else 
  \usepackage{unicode-math}
  \defaultfontfeatures{Scale=MatchLowercase}
  \defaultfontfeatures[\rmfamily]{Ligatures=TeX,Scale=1}
\fi
\IfFileExists{upquote.sty}{\usepackage{upquote}}{}
\IfFileExists{microtype.sty}{
  \usepackage[]{microtype}
  \UseMicrotypeSet[protrusion]{basicmath} 
}{}
\makeatletter
\@ifundefined{KOMAClassName}{
  \IfFileExists{parskip.sty}{%
    \usepackage{parskip}
  }{
    \setlength{\parindent}{0pt}
    \setlength{\parskip}{6pt plus 2pt minus 1pt}}
}{
  \KOMAoptions{parskip=half}}
\makeatother
\usepackage{xcolor}
\IfFileExists{xurl.sty}{\usepackage{xurl}}{} 
\IfFileExists{bookmark.sty}{\usepackage{bookmark}}{\usepackage{hyperref}}
\hypersetup{
  hidelinks,
  pdfcreator={LaTeX via pandoc}}
\usepackage[margin=0.6in]{geometry}
\usepackage{longtable,booktabs}
\usepackage{etoolbox}
\makeatletter
\patchcmd\longtable{\par}{\if@noskipsec\mbox{}\fi\par}{}{}
\makeatother
\IfFileExists{footnotehyper.sty}{\usepackage{footnotehyper}}{\usepackage{footnote}}
\makesavenoteenv{longtable}
\usepackage{graphicx}
\usepackage{float}
\makeatletter
\def\maxwidth{\ifdim\Gin@nat@width>\linewidth\linewidth\else\Gin@nat@width\fi}
\def\maxheight{\ifdim\Gin@nat@height>\textheight\textheight\else\Gin@nat@height\fi}
\makeatother
\setkeys{Gin}{width=\maxwidth,height=\maxheight,keepaspectratio}
\makeatletter
\def\fps@figure{htbp}
\makeatother
\providecommand{\tightlist}{%
  \setlength{\itemsep}{0pt}\setlength{\parskip}{0pt}}
\author{}
\date{}

\begin{document}

\hypertarget{indelfreealigner-a-streaming-aligner-for-comprehensive-gapless-alignment-against-terabase-scale-references}{%
\section{\texorpdfstring{\textbf{IndelFreeAligner: A Streaming Aligner
for Comprehensive Gapless Alignment Against Terabase-Scale
References}}{IndelFreeAligner: A Streaming Aligner for Comprehensive Gapless Alignment Against Terabase-Scale References}}\label{indelfreealigner-a-streaming-aligner-for-comprehensive-gapless-alignment-against-terabase-scale-references}}

Brian Bushnell1*

1DOE Joint Genome Institute, Lawrence Berkeley National Laboratory,
Berkeley, CA, USA

*Corresponding author:
\href{mailto:bbushnell@lbl.gov}{\nolinkurl{bbushnell@lbl.gov}}

\textbf{ORCiD}\\
Brian Bushnell: \url{https://orcid.org/0000-0002-8140-0131}

\hypertarget{abstract}{%
\subsubsection{\texorpdfstring{\textbf{Abstract}}{Abstract}}\label{abstract}}

The comparison of short sequences to massive reference databases is a
cornerstone of modern genomics, but it presents a significant
scalability challenge. Traditional alignment tools rely on time- and
memory-intensive pre-indexing of the reference, creating a substantial
bottleneck for applications involving small query sets against
terabase-scale data, such as CRISPR spacer analysis. Here we present
IndelFreeAligner, a streaming, indel-free alignment tool that eliminates
the preprocessing bottleneck. It operates in two modes: an indexed mode
for larger query sets and a brute-force mode for maximum speed on small
query sets. By processing reference sequences on-the-fly,
IndelFreeAligner supports user-specified mismatch thresholds up to the
full query length and maintains memory usage independent of total
reference size. A novel MinHitsCalculator component uses Monte Carlo
simulation to determine adaptive seed-hit thresholds for indexed mode.
Benchmarks against Bowtie1 and BLAST+ demonstrate that IndelFreeAligner
aligns a single query against a 4 Gbp reference in 1.7 seconds versus 17
minutes for Bowtie1 (including index construction at optimal thread
count), a \textbf{607-fold speedup}. Against RefSeq Bacteria (560 GB
compressed), IndelFreeAligner completes a 10-query search in 12 minutes
using 8 GB of RAM, while BLAST+ requires 3 hours and 17 minutes to build
its database alone and 506 GB of RAM to query it. Indexed mode achieves
0\% false negatives through 4 substitutions (99.84-99.90\% mapped at
8-16); brute-force mode is exhaustive by design. IndelFreeAligner
provides a scalable and efficient solution for alignment tasks that were
previously computationally prohibitive. It is distributed open-source as
part of BBTools (Bushnell, 2014).\\
\textbf{Keywords:} sequence alignment, SIMD optimization, metagenomics,
CRISPR, scalability, streaming algorithms, BBTools

\hypertarget{introduction}{%
\subsubsection{\texorpdfstring{\textbf{Introduction}}{Introduction}}\label{introduction}}

The exponential growth of genomic and metagenomic databases has enabled
biological discovery on an unprecedented scale. However, this data
deluge has also strained the capabilities of foundational bioinformatics
tools. Many alignment algorithms, designed in an era of smaller
reference genomes, face significant challenges when applied to the
terabase-scale datasets arising in modern research. This challenge was
brought into sharp focus by recent landmark studies such as
``Planetary-scale metagenomic search reveals new patterns of CRISPR
targeting'' (Roux et al., 2025). Such ambitious projects, which search
for specific sequences across vast and diverse datasets, push existing
tools to their limits. The study successfully identified novel patterns
but also highlighted a new computational paradigm: the need to align a
relatively small number of query sequences (e.g., CRISPR spacers)
against an ever-expanding universe of reference data. In this context,
the architecture of traditional aligners like Bowtie1 (Langmead et al.,
2009) and BWA (Li \& Durbin, 2009) presents a major bottleneck. These
tools rely on a pre-indexing step (e.g., building an FM-index) that,
while enabling fast queries, is computationally expensive. For a 4 Gbp
reference, Bowtie1's index construction takes 17 minutes at optimal
thread count (16 threads) and scales poorly beyond that. For
terabase-scale references such as RefSeq Bacteria, indexing is estimated
to require days of wall time and terabytes of RAM. BLAST+ (Camacho et
al., 2009) faces similar limitations: building a nucleotide database for
RefSeq Bacteria requires over 3 hours, and querying it demands over 500
GB of RAM. Furthermore, the algorithmic complexity of searching these
index structures often imposes a hard limit on the number of mismatches
that can be tolerated --- Bowtie1, for example, supports at most 3
substitutions. To address this specific computational challenge, we
developed IndelFreeAligner, a streaming alignment tool designed to
eliminate the preprocessing bottleneck. The name refers to its support
for indel-free (Hamming distance) alignments, which most aligners handle
only opportunistically but which are the correct distance metric for
applications where insertions and deletions are not expected.
IndelFreeAligner is optimized for the increasingly common scenario of
aligning a small set of queries against a massive, terabase-scale
reference. It processes reference sequences on-the-fly, requires no
pre-indexing, and uses SIMD-accelerated algorithms to maintain high
performance. It supports user-specified mismatch thresholds with no hard
upper limit, making it ideally suited for the exploratory and
large-scale analyses that are defining the next era of genomics.

\hypertarget{methods}{%
\subsubsection{\texorpdfstring{\textbf{Methods}}{Methods}}\label{methods}}

\hypertarget{algorithm-design}{%
\paragraph{\texorpdfstring{\textbf{Algorithm
Design}}{Algorithm Design}}\label{algorithm-design}}

IndelFreeAligner employs a streaming architecture that processes
reference sequences on-demand. This design choice completely eliminates
the reference preprocessing step, allowing for immediate alignment. The
tool operates in two distinct modes: indexed and brute-force.\\
In \textbf{indexed mode}, query sequences are pre-processed to extract
k-mer seeds using an adaptive multi-K strategy: queries are binned by
length and error rate, and a list of k-mer lengths (default
k=8,9,10,12,14) is evaluated to select the longest k for which
sufficient seed hits are expected for each query. To tolerate
substitutions, a central portion of the k-mer can be masked (e.g.,
XXXXX-NNN-XXXXX). As the reference streams past, each contig is indexed
using a PackedIndex data structure --- a hash-backed compressed sparse
row (CSR) format with singleton optimization and stop-bit encoding that
minimizes memory per k-mer position. A key innovation is the
MinHitsCalculator module, which uses Monte Carlo simulation (default
200,000 iterations, deterministic seed) to determine the minimum number
of seed hits required to identify a valid alignment with a user-defined
probability (default 99.9\%). This threshold is computed per query based
on its valid k-mer count, the allowed substitutions, and the k-mer step
size. Setting the probability to 100\% switches to a deterministic lower
bound on required seed hits (validKmers - kEffective × maxSubs -
maxClips), which guarantees all valid alignments containing at least one
error-free kmer are found, at the cost of additional seed-hit
processing.\\
In \textbf{brute-force mode}, all indexing is bypassed. Instead, the
aligner performs exhaustive alignment testing at every possible start
position using a SIMD-accelerated kernel. This mode guarantees all valid
alignments will be found and reported. It is fastest for very small
query sets, as it incurs zero index-construction overhead but scales
linearly with query count. The crossover with indexed mode is
workload-dependent (see Results).

Both modes support soft-clipped alignments at contig boundaries
(default: up to 25\% of query length), enabling detection of reads that
overhang the ends of reference contigs. Short contigs are fused with
N-padding to enable efficient batch processing; the SIMD kernel tracks
reference Ns separately from substitutions to correctly handle these
fused boundaries.

\hypertarget{simd-optimization}{%
\paragraph{\texorpdfstring{\textbf{SIMD
Optimization}}{SIMD Optimization}}\label{simd-optimization}}

The performance-critical alignment kernel, used in both modes, is
implemented using Java's Vector API to leverage SIMD instructions.
Mismatch counting uses a cascading dual-width strategy: 256-bit vectors
(processing 32 bytes simultaneously), falling back to 64-bit vectors (8
bytes) for remainders, with a scalar tail for the final bytes. This
approach minimizes scalar operations across all sequence lengths. In
indexed mode, the kernel counts mismatches between a query and a single
reference position. In brute-force mode, it processes 32 candidate
positions simultaneously, broadcasting each query base against 32
consecutive reference positions and accumulating per-lane substitution
and clip counts with early termination when all lanes exceed either
threshold.

\hypertarget{memory-management}{%
\paragraph{\texorpdfstring{\textbf{Memory
Management}}{Memory Management}}\label{memory-management}}

IndelFreeAligner's memory usage is proportional to the query count plus
the number of threads times the largest reference contig, independent of
total reference size. For brute-force mode, memory is approximately 1
byte per in-flight base; for indexed mode, 12-26 bytes per in-flight
base due to the per-contig PackedIndex structure. Reference sequences
are processed one at a time per thread, and temporary data structures
are discarded after use. The query set is loaded once at startup. This
design means memory requirements are identical whether the total
reference is 4 Gbp or 400 Gbp --- only the largest individual contig
matters. Both modes of IndelFreeAligner successfully searched all of
RefSeq Bacteria (560 GB compressed) using 8 GB of RAM (brute-force at 64
threads; indexed at 32 threads).

\hypertarget{output-format}{%
\paragraph{\texorpdfstring{\textbf{Output
Format}}{Output Format}}\label{output-format}}

Due to its streaming architecture, IndelFreeAligner does not know the
full set of reference sequences until processing is complete. Primary
SAM output is therefore headerless. A separate header file can be
generated via the \texttt{outh=} flag and concatenated with the
alignment output to produce a standard SAM file.

\hypertarget{benchmarking}{%
\paragraph{\texorpdfstring{\textbf{Benchmarking}}{Benchmarking}}\label{benchmarking}}

Benchmarks were performed on dedicated NERSC Dori cluster nodes with
dual AMD EPYC processors (64 cores, 128 hardware threads) and 2 TB RAM,
running CentOS 7 with Java 21. All alignment tests used 64 threads
unless otherwise noted. Bowtie1 index construction was tested at 1-64
threads; optimal (16 threads) was used for total-time calculations.

Three reference datasets were used: a 4 Mbp bacterial genome
(\emph{Lachnospiraceae} sp., 220 contigs), a 4 Gbp metagenomic assembly
(129,236 contigs), and NCBI RefSeq Bacteria (June 2026 release, 2.17
Tbp, 50.6 million sequences).

For speed benchmarks, simulated 150 bp reads were generated using
RandomReads (BBTools) with substitution errors at a flat Q15 quality
score (\textasciitilde3.2\% per-base error rate). For accuracy
validation, reads with exactly N substitutions (0-16) were generated
using \texttt{snprate=1\ maxsnps=N\ adderrors=f}, producing reads with
known true positions and controlled mismatch counts. Accuracy was
assessed using GradeSam (BBTools), which compares reported alignment
positions against the true positions encoded in the read names.

IndelFreeAligner minimum memory requirements to run successfully were
determined by binary search on the JVM heap size (\texttt{-Xmx}).
Bowtie1 and BLAST+ memory was measured as peak resident set size via
\texttt{/usr/bin/time\ -v}. IFA peak RSS was measured at 64 threads with
\texttt{-Xmx} set 15\% above the empirical minimum.

Software versions: IndelFreeAligner in BBTools 39.99; Bowtie1 1.3.1;
BLAST+ 2.17.0 (Singularity container).

\hypertarget{results}{%
\subsubsection{\texorpdfstring{\textbf{Results}}{Results}}\label{results}}

\hypertarget{reference-preprocessing}{%
\paragraph{\texorpdfstring{\textbf{Reference
Preprocessing}}{Reference Preprocessing}}\label{reference-preprocessing}}

A fundamental difference between IndelFreeAligner and traditional tools
is the elimination of reference preprocessing. Table 1 shows the cost of
this step for Bowtie1 and BLAST+ across three reference scales.

\textbf{Table 1.} Reference preprocessing time and peak memory.
IndelFreeAligner requires no preprocessing.

\begin{longtable}[]{@{}p{3.2cm}p{1.3cm}p{1.8cm}p{1.3cm}p{1.8cm}p{1.3cm}@{}}
\toprule
Reference & Size & Bowtie1 time & Bowtie1 RAM & BLAST+ time & BLAST+
RAM\tabularnewline
\midrule
\endhead
Bacterial genome & 4 Mbp & 1.6 s & 99 MB & 0.6 s & 48 MB\tabularnewline
Metagenomic assembly & 4 Gbp & 1,031 s & 8.2 GB & 22 s & 47
MB\tabularnewline
RefSeq Bacteria & 2.17 Tbp & not run & --- & 11,801 s & 736
MB\tabularnewline
\bottomrule
\end{longtable}

Bowtie1's index construction was benchmarked at thread counts from 1 to
64 (Table 2). Optimal performance on the 4 Gbp reference was achieved at
16 threads (17:11); performance degraded at 32 threads (19:02) and 64
threads (22:57). Single-threaded construction required 49:42. Bowtie1
supports large references via its \texttt{-\/-large-index} mode but
index construction at the RefSeq Bacteria scale is estimated to require
days of wall time and terabytes of RAM. BLAST+'s makeblastdb is
single-threaded; RefSeq Bacteria was streamed to makeblastdb via pipe.

\newpage
\textbf{Table 2.} Bowtie1 bowtie-build wall-clock time vs.~thread count
on the 4 Gbp reference.

\begin{longtable}[]{@{}llll@{}}
\toprule
Threads & Time & Speedup & Peak RAM\tabularnewline
\midrule
\endhead
1 & 49 min 42 s & 1.0x & 7.3 GB\tabularnewline
2 & 30 min 00 s & 1.7x & 8.8 GB\tabularnewline
4 & 22 min 38 s & 2.2x & 8.1 GB\tabularnewline
8 & 18 min 06 s & 2.7x & 8.2 GB\tabularnewline
16 & 17 min 11 s & 2.9x & 8.2 GB\tabularnewline
32 & 19 min 02 s & 2.6x & 8.7 GB\tabularnewline
64 & 22 min 57 s & 2.2x & 9.9 GB\tabularnewline
\bottomrule
\end{longtable}

\hypertarget{large-scale-reference-performance-4-gbp-metagenome}{%
\paragraph{\texorpdfstring{\textbf{Large-Scale Reference Performance (4
Gbp
metagenome)}}{Large-Scale Reference Performance (4 Gbp metagenome)}}\label{large-scale-reference-performance-4-gbp-metagenome}}

The primary design goal of IndelFreeAligner is to excel at aligning few
queries to a large reference. Table 3 and Figure 1A show total time
(including all preprocessing) and alignment-only time for each tool.

\textbf{Table 3.} Wall-clock time (seconds) for aligning against the 4
Gbp metagenomic reference at 3 substitutions. Total time includes
index/database construction at optimal thread count. Alignment-only time
is shown in parentheses.

\begin{longtable}[]{@{}lllll@{}}
\toprule
\begin{minipage}[b]{0.09\columnwidth}\raggedright
Queries\strut
\end{minipage} & \begin{minipage}[b]{0.15\columnwidth}\raggedright
IFA brute-force\strut
\end{minipage} & \begin{minipage}[b]{0.12\columnwidth}\raggedright
IFA indexed\strut
\end{minipage} & \begin{minipage}[b]{0.25\columnwidth}\raggedright
Bowtie1 total (align only)\strut
\end{minipage} & \begin{minipage}[b]{0.25\columnwidth}\raggedright
BLAST+ total (align only)\strut
\end{minipage}\tabularnewline
\midrule
\endhead
\begin{minipage}[t]{0.09\columnwidth}\raggedright
1\strut
\end{minipage} & \begin{minipage}[t]{0.15\columnwidth}\raggedright
\textbf{1.7}\strut
\end{minipage} & \begin{minipage}[t]{0.12\columnwidth}\raggedright
12\strut
\end{minipage} & \begin{minipage}[t]{0.25\columnwidth}\raggedright
1,032 (1.3)\strut
\end{minipage} & \begin{minipage}[t]{0.25\columnwidth}\raggedright
23 (0.7)\strut
\end{minipage}\tabularnewline
\begin{minipage}[t]{0.09\columnwidth}\raggedright
10\strut
\end{minipage} & \begin{minipage}[t]{0.15\columnwidth}\raggedright
\textbf{6.3}\strut
\end{minipage} & \begin{minipage}[t]{0.12\columnwidth}\raggedright
13\strut
\end{minipage} & \begin{minipage}[t]{0.25\columnwidth}\raggedright
1,032 (1.2)\strut
\end{minipage} & \begin{minipage}[t]{0.25\columnwidth}\raggedright
24 (1.9)\strut
\end{minipage}\tabularnewline
\begin{minipage}[t]{0.09\columnwidth}\raggedright
100\strut
\end{minipage} & \begin{minipage}[t]{0.15\columnwidth}\raggedright
112\strut
\end{minipage} & \begin{minipage}[t]{0.12\columnwidth}\raggedright
\textbf{12}\strut
\end{minipage} & \begin{minipage}[t]{0.25\columnwidth}\raggedright
1,032 (1.2)\strut
\end{minipage} & \begin{minipage}[t]{0.25\columnwidth}\raggedright
36 (14)\strut
\end{minipage}\tabularnewline
\begin{minipage}[t]{0.09\columnwidth}\raggedright
1,000\strut
\end{minipage} & \begin{minipage}[t]{0.15\columnwidth}\raggedright
1,074\strut
\end{minipage} & \begin{minipage}[t]{0.12\columnwidth}\raggedright
\textbf{13}\strut
\end{minipage} & \begin{minipage}[t]{0.25\columnwidth}\raggedright
1,032 (1.2)\strut
\end{minipage} & \begin{minipage}[t]{0.25\columnwidth}\raggedright
143 (121)\strut
\end{minipage}\tabularnewline
\begin{minipage}[t]{0.09\columnwidth}\raggedright
10,000\strut
\end{minipage} & \begin{minipage}[t]{0.15\columnwidth}\raggedright
11,558\strut
\end{minipage} & \begin{minipage}[t]{0.12\columnwidth}\raggedright
\textbf{15}\strut
\end{minipage} & \begin{minipage}[t]{0.25\columnwidth}\raggedright
1,032 (1.3)\strut
\end{minipage} & \begin{minipage}[t]{0.25\columnwidth}\raggedright
1,220 (1,198)\strut
\end{minipage}\tabularnewline
\begin{minipage}[t]{0.09\columnwidth}\raggedright
100,000\strut
\end{minipage} & \begin{minipage}[t]{0.15\columnwidth}\raggedright
---\strut
\end{minipage} & \begin{minipage}[t]{0.12\columnwidth}\raggedright
\textbf{34}\strut
\end{minipage} & \begin{minipage}[t]{0.25\columnwidth}\raggedright
1,033 (1.5)\strut
\end{minipage} & \begin{minipage}[t]{0.25\columnwidth}\raggedright
---\strut
\end{minipage}\tabularnewline
\begin{minipage}[t]{0.09\columnwidth}\raggedright
1,000,000\strut
\end{minipage} & \begin{minipage}[t]{0.15\columnwidth}\raggedright
---\strut
\end{minipage} & \begin{minipage}[t]{0.12\columnwidth}\raggedright
\textbf{214}\strut
\end{minipage} & \begin{minipage}[t]{0.25\columnwidth}\raggedright
1,038 (6.5)\strut
\end{minipage} & \begin{minipage}[t]{0.25\columnwidth}\raggedright
---\strut
\end{minipage}\tabularnewline
\bottomrule
\end{longtable}

IndelFreeAligner's brute-force mode aligned a single query in 1.7
seconds --- a \textbf{607-fold speedup} over Bowtie1's total time. Its
indexed mode completed all query set sizes through 100,000 queries in 34
seconds or less, still \textbf{30-fold faster} than Bowtie1 at 100,000
queries. The brute-force mode was fastest for 1-10 queries (1.7 s for a
single query) but scales linearly and becomes impractical above
\textasciitilde100 queries at this reference size.

When considering alignment time only (excluding preprocessing), Bowtie1
is generally fastest; BLAST+ was faster for a single query and
competitive at small query counts. This underscores that
IndelFreeAligner's advantage is architectural: by eliminating
preprocessing, it converts a 17-minute fixed cost into zero, which
dominates total runtime whenever the query set is small relative to the
reference.

\begin{figure}[H]
\centering
\includegraphics[width=1\textwidth,height=\textheight]{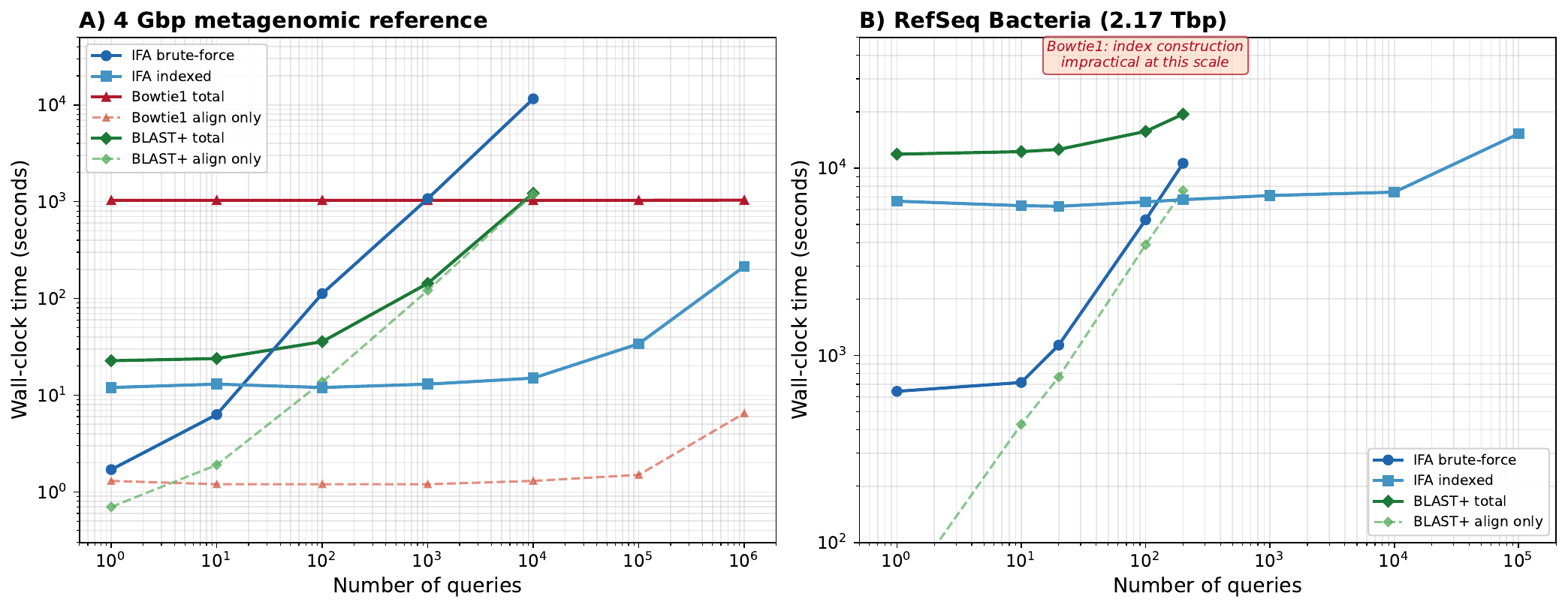}
\caption{Total wall-clock time vs.~query count. (A) 4 Gbp metagenomic
reference. (B) RefSeq Bacteria (2.17 Tbp). Solid lines include
preprocessing; dashed lines show alignment time only. IndelFreeAligner
requires no preprocessing at any scale.}
\end{figure}

\hypertarget{small-scale-reference-performance-4-mbp-bacterial-genome}{%
\paragraph{\texorpdfstring{\textbf{Small-Scale Reference Performance (4
Mbp bacterial
genome)}}{Small-Scale Reference Performance (4 Mbp bacterial genome)}}\label{small-scale-reference-performance-4-mbp-bacterial-genome}}

In the reverse scenario --- aligning against a 4 Mbp bacterial genome
--- Bowtie1's sub-2-second indexing and highly optimized query engine
gave it a substantial performance advantage for large query sets (Table
4).

\textbf{Table 4.} Wall-clock time (seconds) for aligning against the 4
Mbp bacterial reference at 3 substitutions.

\begin{longtable}[]{@{}lllll@{}}
\toprule
\begin{minipage}[b]{0.09\columnwidth}\raggedright
Queries\strut
\end{minipage} & \begin{minipage}[b]{0.15\columnwidth}\raggedright
IFA brute-force\strut
\end{minipage} & \begin{minipage}[b]{0.12\columnwidth}\raggedright
IFA indexed\strut
\end{minipage} & \begin{minipage}[b]{0.25\columnwidth}\raggedright
Bowtie1 total (align only)\strut
\end{minipage} & \begin{minipage}[b]{0.25\columnwidth}\raggedright
BLAST+ total (align only)\strut
\end{minipage}\tabularnewline
\midrule
\endhead
\begin{minipage}[t]{0.09\columnwidth}\raggedright
1\strut
\end{minipage} & \begin{minipage}[t]{0.15\columnwidth}\raggedright
\textbf{0.66}\strut
\end{minipage} & \begin{minipage}[t]{0.12\columnwidth}\raggedright
0.69\strut
\end{minipage} & \begin{minipage}[t]{0.25\columnwidth}\raggedright
2.8 (1.2)\strut
\end{minipage} & \begin{minipage}[t]{0.25\columnwidth}\raggedright
1.3 (0.74)\strut
\end{minipage}\tabularnewline
\begin{minipage}[t]{0.09\columnwidth}\raggedright
10\strut
\end{minipage} & \begin{minipage}[t]{0.15\columnwidth}\raggedright
0.55\strut
\end{minipage} & \begin{minipage}[t]{0.12\columnwidth}\raggedright
\textbf{0.45}\strut
\end{minipage} & \begin{minipage}[t]{0.25\columnwidth}\raggedright
2.2 (0.66)\strut
\end{minipage} & \begin{minipage}[t]{0.25\columnwidth}\raggedright
1.3 (0.68)\strut
\end{minipage}\tabularnewline
\begin{minipage}[t]{0.09\columnwidth}\raggedright
100\strut
\end{minipage} & \begin{minipage}[t]{0.15\columnwidth}\raggedright
0.82\strut
\end{minipage} & \begin{minipage}[t]{0.12\columnwidth}\raggedright
\textbf{0.68}\strut
\end{minipage} & \begin{minipage}[t]{0.25\columnwidth}\raggedright
2.2 (0.66)\strut
\end{minipage} & \begin{minipage}[t]{0.25\columnwidth}\raggedright
3.0 (2.3)\strut
\end{minipage}\tabularnewline
\begin{minipage}[t]{0.09\columnwidth}\raggedright
1,000\strut
\end{minipage} & \begin{minipage}[t]{0.15\columnwidth}\raggedright
3.1\strut
\end{minipage} & \begin{minipage}[t]{0.12\columnwidth}\raggedright
\textbf{0.59}\strut
\end{minipage} & \begin{minipage}[t]{0.25\columnwidth}\raggedright
2.2 (0.66)\strut
\end{minipage} & \begin{minipage}[t]{0.25\columnwidth}\raggedright
19 (18)\strut
\end{minipage}\tabularnewline
\begin{minipage}[t]{0.09\columnwidth}\raggedright
10,000\strut
\end{minipage} & \begin{minipage}[t]{0.15\columnwidth}\raggedright
26\strut
\end{minipage} & \begin{minipage}[t]{0.12\columnwidth}\raggedright
\textbf{0.57}\strut
\end{minipage} & \begin{minipage}[t]{0.25\columnwidth}\raggedright
2.2 (0.66)\strut
\end{minipage} & \begin{minipage}[t]{0.25\columnwidth}\raggedright
73 (72)\strut
\end{minipage}\tabularnewline
\begin{minipage}[t]{0.09\columnwidth}\raggedright
100,000\strut
\end{minipage} & \begin{minipage}[t]{0.15\columnwidth}\raggedright
252\strut
\end{minipage} & \begin{minipage}[t]{0.12\columnwidth}\raggedright
\textbf{1.0}\strut
\end{minipage} & \begin{minipage}[t]{0.25\columnwidth}\raggedright
2.2 (0.67)\strut
\end{minipage} & \begin{minipage}[t]{0.25\columnwidth}\raggedright
594 (594)\strut
\end{minipage}\tabularnewline
\begin{minipage}[t]{0.09\columnwidth}\raggedright
1,000,000\strut
\end{minipage} & \begin{minipage}[t]{0.15\columnwidth}\raggedright
2,499\strut
\end{minipage} & \begin{minipage}[t]{0.12\columnwidth}\raggedright
3.9\strut
\end{minipage} & \begin{minipage}[t]{0.25\columnwidth}\raggedright
\textbf{3.7 (2.2)}\strut
\end{minipage} & \begin{minipage}[t]{0.25\columnwidth}\raggedright
---\strut
\end{minipage}\tabularnewline
\begin{minipage}[t]{0.09\columnwidth}\raggedright
10,000,000\strut
\end{minipage} & \begin{minipage}[t]{0.15\columnwidth}\raggedright
---\strut
\end{minipage} & \begin{minipage}[t]{0.12\columnwidth}\raggedright
31\strut
\end{minipage} & \begin{minipage}[t]{0.25\columnwidth}\raggedright
\textbf{20 (19)}\strut
\end{minipage} & \begin{minipage}[t]{0.25\columnwidth}\raggedright
---\strut
\end{minipage}\tabularnewline
\bottomrule
\end{longtable}

IndelFreeAligner's indexed mode remained faster than Bowtie1 up to
approximately 100,000 queries; at 1 million queries Bowtie1 was slightly
faster (3.7 s vs 3.9 s), and at 10 million queries Bowtie1 was 1.5x
faster (20 s vs 31 s), demonstrating that traditional indexed aligners
retain an advantage when the reference is small enough to index cheaply
and the query set is very large. BLAST+ was consistently slower than
both other tools at query counts beyond 10 on the small reference.

\hypertarget{terabase-scale-performance-refseq-bacteria}{%
\paragraph{\texorpdfstring{\textbf{Terabase-Scale Performance (RefSeq
Bacteria)}}{Terabase-Scale Performance (RefSeq Bacteria)}}\label{terabase-scale-performance-refseq-bacteria}}

To demonstrate IndelFreeAligner's capabilities at scales where
traditional tools become impractical, we aligned queries against the
complete NCBI RefSeq Bacteria collection (2.17 Tbp, June 2026 release).
Bowtie1 supports large references via \texttt{-\/-large-index}, but
linearly scaling from its 4 Gbp performance (Table 2: 1,031 s and 8.2 GB
at 4 Gbp), index construction at 2.17 Tbp is estimated to require days
of wall time and terabytes of RAM; this was not tested. BLAST+ required
3 hours 17 minutes to build its database and 506 GB of RAM to load it
for queries.

\newpage
\textbf{Table 5.} Wall-clock time for aligning against RefSeq Bacteria.
Separate minimum-heap tests confirmed both IFA modes run within 8 GB of
RAM.

\begin{longtable}[]{@{}llll@{}}
\toprule
Queries & IFA brute-force & IFA indexed (32 threads) & BLAST+ total
(query only)\tabularnewline
\midrule
\endhead
1 & \textbf{10 min 42 s} & 1 h 51 min & 3 h 18 min (49 s)\tabularnewline
10 & \textbf{11 min 56 s} & 1 h 45 min & 3 h 24 min (7
min)\tabularnewline
20 & \textbf{18 min 54 s} & 1 h 44 min & 3 h 30 min (13
min)\tabularnewline
100 & \textbf{1 h 28 min} & 1 h 50 min & 4 h 22 min (1 h 5
min)\tabularnewline
200 & 2 h 57 min & \textbf{1 h 53 min} & 5 h 24 min (2 h 7
min)\tabularnewline
1,000 & --- & \textbf{1 h 59 min} & ---\tabularnewline
10,000 & --- & \textbf{2 h 4 min} & ---\tabularnewline
100,000 & --- & \textbf{4 h 14 min} & ---\tabularnewline
\bottomrule
\end{longtable}

IndelFreeAligner's brute-force mode completed a single-query search
against the entire RefSeq Bacteria collection in under 11 minutes using
8 GB of RAM, making it feasible to screen individual sequences (e.g.,
CRISPR spacers, pathogen markers) against the full breadth of known
bacterial diversity within an 8 GB memory budget. Brute-force mode was
faster than indexed mode through 100 queries; the crossover occurs
between 100 and 200 queries. Indexed mode time was dominated by
reference streaming and per-contig index construction, growing only
modestly from 1 hour 45 minutes at 10 queries to 2 hours 4 minutes at
10,000 queries. Indexed mode required 32 threads to fit within 8 GB;
brute-force mode ran at 64 threads within the same memory budget.

\hypertarget{accuracy-validation}{%
\paragraph{\texorpdfstring{\textbf{Accuracy
Validation}}{Accuracy Validation}}\label{accuracy-validation}}

To validate alignment correctness, we generated 10,000 reads with
exactly N substitutions (N = 0, 1, 2, 3, 4, 8, 16) from the 4 Gbp
metagenomic reference and aligned them with each tool at the
corresponding mismatch threshold. Accuracy was assessed using GradeSam
(BBTools), which compares reported primary alignment positions against
the true positions encoded in the read names.

\textbf{Table 6.} Alignment accuracy on the 4 Gbp metagenomic reference
(10,000 reads per mismatch level, 150 bp). True-origin primary: primary
alignment placed at the read's true origin. Alternate-valid primary:
primary placed at an equally valid duplicate position (true origin
present as secondary). Unmapped: read not reported. Bowtie1 limited to
subs \textless= 3.

\begin{longtable}[]{@{}llll@{}}
\toprule
\begin{minipage}[b]{0.05\columnwidth}\raggedright
Subs\strut
\end{minipage} & \begin{minipage}[b]{0.25\columnwidth}\raggedright
Bowtie1 origin / alt / unmapped\strut
\end{minipage} & \begin{minipage}[b]{0.28\columnwidth}\raggedright
IFA indexed origin / alt / unmapped\strut
\end{minipage} & \begin{minipage}[b]{0.31\columnwidth}\raggedright
IFA brute-force origin / alt / unmapped\strut
\end{minipage}\tabularnewline
\midrule
\endhead
\begin{minipage}[t]{0.05\columnwidth}\raggedright
0\strut
\end{minipage} & \begin{minipage}[t]{0.25\columnwidth}\raggedright
99.99 / 0.01 / 0\strut
\end{minipage} & \begin{minipage}[t]{0.28\columnwidth}\raggedright
99.94 / 0.06 / 0\strut
\end{minipage} & \begin{minipage}[t]{0.31\columnwidth}\raggedright
99.95 / 0.05 / 0\strut
\end{minipage}\tabularnewline
\begin{minipage}[t]{0.05\columnwidth}\raggedright
1\strut
\end{minipage} & \begin{minipage}[t]{0.25\columnwidth}\raggedright
99.98 / 0.02 / 0\strut
\end{minipage} & \begin{minipage}[t]{0.28\columnwidth}\raggedright
99.95 / 0.05 / 0\strut
\end{minipage} & \begin{minipage}[t]{0.31\columnwidth}\raggedright
99.95 / 0.05 / 0\strut
\end{minipage}\tabularnewline
\begin{minipage}[t]{0.05\columnwidth}\raggedright
2\strut
\end{minipage} & \begin{minipage}[t]{0.25\columnwidth}\raggedright
99.98 / 0.02 / 0\strut
\end{minipage} & \begin{minipage}[t]{0.28\columnwidth}\raggedright
99.97 / 0.03 / 0\strut
\end{minipage} & \begin{minipage}[t]{0.31\columnwidth}\raggedright
99.95 / 0.05 / 0\strut
\end{minipage}\tabularnewline
\begin{minipage}[t]{0.05\columnwidth}\raggedright
3\strut
\end{minipage} & \begin{minipage}[t]{0.25\columnwidth}\raggedright
99.99 / 0.01 / 0\strut
\end{minipage} & \begin{minipage}[t]{0.28\columnwidth}\raggedright
99.97 / 0.03 / 0\strut
\end{minipage} & \begin{minipage}[t]{0.31\columnwidth}\raggedright
99.95 / 0.05 / 0\strut
\end{minipage}\tabularnewline
\begin{minipage}[t]{0.05\columnwidth}\raggedright
4\strut
\end{minipage} & \begin{minipage}[t]{0.25\columnwidth}\raggedright
---\strut
\end{minipage} & \begin{minipage}[t]{0.28\columnwidth}\raggedright
99.95 / 0.05 / 0\strut
\end{minipage} & \begin{minipage}[t]{0.31\columnwidth}\raggedright
99.92 / 0.08 / 0\strut
\end{minipage}\tabularnewline
\begin{minipage}[t]{0.05\columnwidth}\raggedright
8\strut
\end{minipage} & \begin{minipage}[t]{0.25\columnwidth}\raggedright
---\strut
\end{minipage} & \begin{minipage}[t]{0.28\columnwidth}\raggedright
99.85 / 0.05 / 0.10\strut
\end{minipage} & \begin{minipage}[t]{0.31\columnwidth}\raggedright
99.94 / 0.06 / 0\strut
\end{minipage}\tabularnewline
\begin{minipage}[t]{0.05\columnwidth}\raggedright
16\strut
\end{minipage} & \begin{minipage}[t]{0.25\columnwidth}\raggedright
---\strut
\end{minipage} & \begin{minipage}[t]{0.28\columnwidth}\raggedright
99.74 / 0.10 / 0.16\strut
\end{minipage} & \begin{minipage}[t]{0.31\columnwidth}\raggedright
99.91 / 0.09 / 0\strut
\end{minipage}\tabularnewline
\bottomrule
\end{longtable}

All three tools achieved 0\% false negatives through their supported
mismatch ranges: Bowtie1 at subs 0-3 (its maximum), and IFA indexed mode
at subs 0-4. At higher mismatch levels (subs 8, 16), IFA indexed mode
showed small false negative rates (0.10\% and 0.16\%), consistent with
the 99.9\% probabilistic seed-hit threshold allowing rare misses when
the number of valid k-mers drops at high substitution rates. IFA
brute-force mode is exhaustive and guaranteed to find all valid
alignments; empirical testing confirmed 0\% FN (100\% mapped) at every
level tested (subs 0 through 16).

Minor differences in primary placement (0.01-0.10\% of mapped reads)
reflect primary alignment selection among equally valid positions in
duplicated reference regions, not incorrect alignments. Because
IndelFreeAligner streams the reference rather than reads, the primary is
the first valid alignment encountered in reference order; the true
origin is always present in the reported alignment set (primary or
secondary). IFA indexed mode additionally reports valid soft-clipped
alignments at contig boundaries (up to 25\% overhang) that Bowtie1 does
not find, as Bowtie1 requires full-length alignment.

\hypertarget{mismatch-tolerance-scalability}{%
\paragraph{\texorpdfstring{\textbf{Mismatch Tolerance
Scalability}}{Mismatch Tolerance Scalability}}\label{mismatch-tolerance-scalability}}

Unlike Bowtie1, which is limited to a maximum of 3 substitutions,
IndelFreeAligner supports user-specified mismatch thresholds up to the
full query length. Figure 2 shows IFA indexed mode runtime versus
allowed substitutions on the 4 Gbp reference with 100,000 reads. Runtime
increased only 2.4x from subs=0 (31 s) to subs=16 (76 s), with near-flat
performance through subs=3. This stability arises because indexed mode
runtime at low substitution counts is dominated by per-contig index
construction rather than alignment verification. In brute-force mode,
runtime scales linearly with allowed substitutions due to the SIMD
kernel's early-exit mechanism.

\begin{figure}[H]
\centering
\includegraphics[width=0.7\textwidth,height=\textheight]{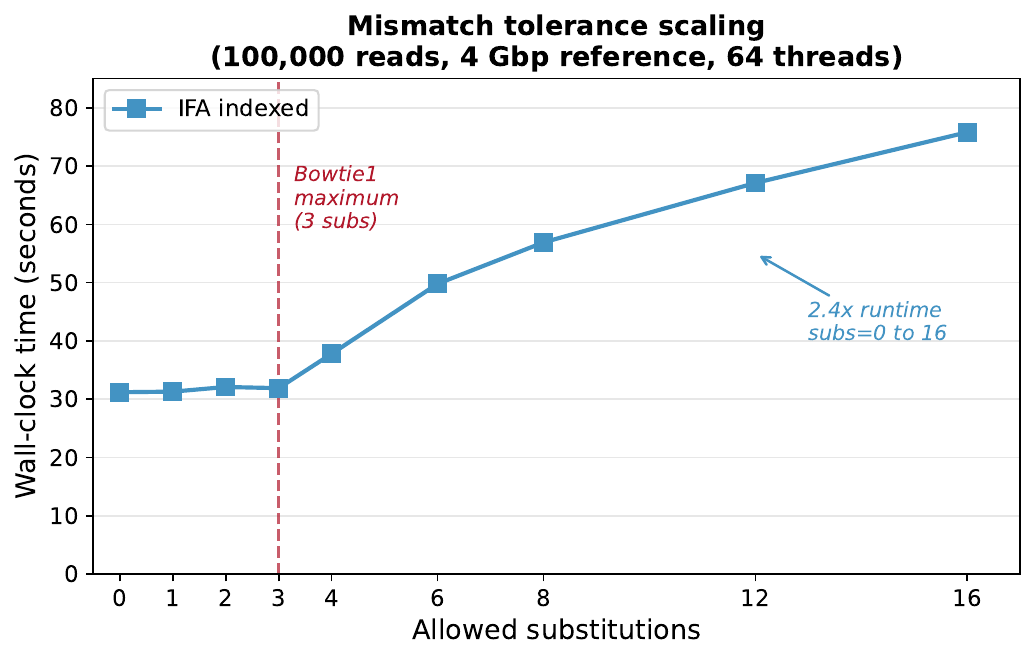}
\caption{IFA indexed mode wall-clock time vs.~allowed substitutions
(100,000 reads, 4 Gbp reference, 64 threads). Bowtie1's maximum of 3
substitutions is shown as a dashed line. Performance increases only 2.4x
from subs=0 to subs=16.}
\end{figure}

\hypertarget{memory-usage}{%
\paragraph{\texorpdfstring{\textbf{Memory
Usage}}{Memory Usage}}\label{memory-usage}}

\textbf{Table 7.} Peak memory for each tool on the 4 Gbp metagenomic
reference (3 substitutions, 64 threads). IFA values are peak RSS
measured with \texttt{-Xmx} set 15\% above the empirical minimum heap.

\begin{longtable}[]{@{}llll@{}}
\toprule
Tool & 100 queries & 10,000 queries & 100,000 queries\tabularnewline
\midrule
\endhead
IFA indexed & 3.4 GB & 3.4 GB & 3.5 GB\tabularnewline
IFA brute-force & 0.8 GB & --- & ---\tabularnewline
Bowtie1 & 3.3 GB & 3.4 GB & 3.6 GB\tabularnewline
BLAST+ & 2.2 GB & 2.9 GB & ---\tabularnewline
\bottomrule
\end{longtable}

On the 4 Gbp reference, IFA indexed mode and Bowtie1 used comparable
memory (\textasciitilde3.4 GB). IFA brute-force mode used about
one-quarter the memory (0.8 GB) because it does not build per-contig
indexes. The key architectural difference is that IndelFreeAligner's
memory is bounded by the largest contig size (times the number of
threads), not total reference size. Both IFA modes searched all of
RefSeq Bacteria (560 GB compressed) within 8 GB of RAM. By contrast,
BLAST+ required 506 GB to load its RefSeq Bacteria database, exceeding
the RAM available on most systems.

\hypertarget{discussion}{%
\subsubsection{\texorpdfstring{\textbf{Discussion}}{Discussion}}\label{discussion}}

IndelFreeAligner is a direct response to a new computational paradigm in
genomics: the need to search a small number of queries against large,
rapidly-growing reference databases. By adopting a streaming
architecture with two distinct operational modes, it inverts the
traditional performance trade-off.

For exploratory analysis with one or a few queries, the brute-force mode
offers the lowest total time among tools tested by eliminating all
preprocessing and all index-construction overhead. For larger query
sets, the indexed mode amortizes the cost of per-contig index
construction across many queries while maintaining the streaming
advantage. The crossover between modes depends on reference size: within
noise for a 4 Mbp reference until indexed mode pulls ahead at 100-1,000
queries, and between 100 and 200 queries for RefSeq Bacteria. Users can
select the mode explicitly or allow the tool to choose automatically
based on query count.

The three-way comparison with Bowtie1 and BLAST+ reveals that
IndelFreeAligner's advantage is architectural rather than algorithmic in
the narrow sense. When alignment-only time is considered (excluding
preprocessing), Bowtie1's FM-index search is extremely fast, and
BLAST+'s heuristic search is competitive for small query counts.
IndelFreeAligner's contribution is eliminating the fixed preprocessing
cost that dominates total runtime in the small-query, large-reference
regime, as well as the memory footprint of monolithic indexes.

It is important to note that IndelFreeAligner is purpose-built for
end-to-end gapless alignment. It is not a replacement for
general-purpose gapped aligners like BWA-MEM (Li, 2013) for applications
such as variant calling or RNA-seq. However, for applications where
substitutions are the primary mode of variation --- including exact or
near-exact sequence matching across large databases --- its performance
and scalability offer a significant advantage.

\hypertarget{data-and-code-availability}{%
\subsubsection{\texorpdfstring{\textbf{Data and Code
Availability}}{Data and Code Availability}}\label{data-and-code-availability}}

IndelFreeAligner is distributed with BBTools 39.99, available at:
\url{https://bbmap.org} Bowtie1 version 1.3.1:
\url{https://bowtie-bio.sourceforge.net/index.shtml}\\
BLAST+ version 2.17.0:
\url{https://blast.ncbi.nlm.nih.gov/doc/blast-help/downloadblastdata.html}\\
Benchmark scripts and figure generation code are available at
https://github.com/bbushnell/IndelFreeAligner-benchmarks.

Example command lines:

{\small\begin{verbatim}
# Read simulation (controlled substitutions)
randomreads.sh in=ref.fa out=reads.fq.gz reads=10000 \
  snprate=1 maxsnps=3 adderrors=f \
  mininsert=150 maxinsert=150 paired=f

# Read simulation (quality-based errors)
randomreads.sh in=ref.fa out=reads.fq.gz \
  reads=100000 adderrors q=15 paired=f

# Bowtie1
bowtie-build --threads 16 ref.fa ref
bowtie -a -v 3 -p 64 -S \
  -x ref -q reads.fq.gz output.sam

# BLAST+ (ungapped, short queries)
reformat.sh in=reads.fq.gz out=reads.fa
makeblastdb -in ref.fa -dbtype nucl -out ref_db
blastn -task blastn-short -ungapped \
  -query reads.fa -db ref_db -num_threads 64 \
  -outfmt 6 -evalue 1000 -word_size 7 \
  -dust no -out output.txt

# IndelFreeAligner (indexed and brute-force)
indelfree.sh ref=ref.fa subs=3 \
  in=reads.fq.gz out=output.sam t=64
indelfree.sh ref=ref.fa subs=3 \
  in=reads.fq.gz out=output.sam index=f t=64

# Accuracy grading
gradesam.sh in=output.sam reads=10000
\end{verbatim}}

\hypertarget{conclusion}{%
\subsubsection{\texorpdfstring{\textbf{Conclusion}}{Conclusion}}\label{conclusion}}

IndelFreeAligner provides a scalable, efficient, and versatile solution
for end-to-end gapless alignment against terabase-scale references. By
bypassing the massive memory and time overheads associated with static
index construction, it seamlessly scales to databases of any size while
maintaining a constant memory footprint relative to reference size. Its
flexible mismatch tolerance --- uncoupled from the hard limits of
traditional index structures --- and robust accuracy provide a powerful,
accessible framework for the next generation of large-scale genomic
discovery.

\hypertarget{acknowledgments}{%
\subsubsection{\texorpdfstring{\textbf{Acknowledgments}}{Acknowledgments}}\label{acknowledgments}}

The work conducted by the U.S. Department of Energy Joint Genome
Institute (\url{https://ror.org/04xm1d337}), a DOE Office of Science
User Facility, is supported by the Office of Science of the U.S.
Department of Energy operated under Contract No.~DE-AC02-05CH11231.

The author thanks Eru for his extensive support in writing this
manuscript. AI tools (Anthropic Claude and OpenAI ChatGPT) were used for
assistance with benchmarking, figure generation, manuscript preparation,
and review.

\hypertarget{references}{%
\subsubsection{\texorpdfstring{\textbf{References}}{References}}\label{references}}

\begin{enumerate}
\def\labelenumi{\arabic{enumi}.}
\tightlist
\item
  Langmead, B., Trapnell, C., Pop, M., \& Salzberg, S. L. (2009).
  Ultrafast and memory-efficient alignment of short DNA sequences to the
  human genome. \emph{Genome Biology}, 10(3), R25.\\
\item
  Li, H., \& Durbin, R. (2009). Fast and accurate short read alignment
  with Burrows-Wheeler transform. \emph{Bioinformatics}, 25(14),
  1754-1760.\\
\item
  Li, H. (2013). Aligning sequence reads, clone sequences and assembly
  contigs with BWA-MEM. \emph{arXiv preprint arXiv:1303.3997}.\\
\item
  Camacho, C., Coulouris, G., Avagyan, V., et al.~(2009). BLAST+:
  architecture and applications. \emph{BMC Bioinformatics}, 10, 421.
\item
  Bushnell, B. (2014). \emph{BBMap: a fast, accurate, splice-aware
  aligner}. \url{https://www.osti.gov/servlets/purl/1241166}.
\item
  Roux, S., Neri, U., Bushnell, B., Fremin, B., George, N. A., Gophna,
  U., Hug, L. A., Camargo, A. P., Wu, D., Ivanova, N., Kyrpides, N., \&
  Eloe-Fadrosh, E. (2025). Planetary-scale metagenomic search reveals
  new patterns of CRISPR targeting. \emph{bioRxiv}.
  doi:10.1101/2025.06.12.659409.
\end{enumerate}

\end{document}